\documentclass[prl,twocolumn,amsmath,amssymb,superscriptaddress]{revtex4}
\usepackage{graphicx}
\usepackage{dcolumn}
\usepackage{bm}

\begin{document}



\title{Gate tunable infrared phonon anomalies in bilayer graphene}

\author{A. B. Kuzmenko}
\affiliation{D\'{e}partement de Physique de la Mati\`{e}re
Condens\'{e}e, Universit\'{e} de Gen\`{e}ve, CH-1211 Gen\`{e}ve 4,
Switzerland}

\author{L. Benfatto}
\affiliation{Centro Studi e Ricerche ``Enrico Fermi'', via
Panisperna 89/A, I-00184, Rome, Italy}\affiliation{SMC Research
Center, CNR-INFM, and ISC-CNR, via dei Taurini 19, 00185 Roma,
Italy} \affiliation{Dipart. di Fisica, Universit\`a ``La
Sapienza'', P.le A. Moro 2, 00185 Rome, Italy}

\author{E. Cappelluti}
\affiliation{SMC Research Center, CNR-INFM, and ISC-CNR, via
dei Taurini 19, 00185 Roma, Italy} \affiliation{Dipart. di
Fisica, Universit\`a ``La Sapienza'', P.le A. Moro 2, 00185
Rome, Italy}
\author{I. Crassee}
\affiliation{D\'{e}partement de Physique de la Mati\`{e}re
Condens\'{e}e, Universit\'{e} de Gen\`{e}ve, CH-1211 Gen\`{e}ve 4,
Switzerland}

\author{D. van der Marel}
\affiliation{D\'{e}partement de Physique de la Mati\`{e}re
Condens\'{e}e, Universit\'{e} de Gen\`{e}ve, CH-1211 Gen\`{e}ve 4,
Switzerland}

\author{P. Blake}
\affiliation{Manchester Centre for Mesoscience and
Nanotechnology, University of Manchester, Manchester M13 9PL,
UK }

\author{K. S. Novoselov}
\affiliation{Manchester Centre for Mesoscience and
Nanotechnology, University of Manchester, Manchester M13 9PL,
UK }

\author{A. K. Geim}
\affiliation{Manchester Centre for Mesoscience and
Nanotechnology, University of Manchester, Manchester M13 9PL,
UK }

\begin{abstract}
We observe a giant increase of the infrared intensity and a
softening of the in-plane antisymmetric phonon mode $E_u$
($\sim$ 0.2 eV) in bilayer graphene as a function of the
gate-induced doping. The phonon peak has a pronounced Fano-like
asymmetry. We suggest that the intensity growth and the
softening originate from the coupling of the phonon mode to the
narrow electronic transition between parallel bands of the same
character, while the asymmetry is due to the interaction with
the continuum of transitions between the lowest hole and
electron bands. The growth of the peak can be interpreted as a
"charged-phonon" effect observed previously in organic chain
conductors and doped fullerenes, which can be tuned in graphene
with the gate voltage.
\end{abstract}

\maketitle

There is a lot of interest in electronic properties of
graphene, especially, due to the chiral nature of its
quasiparticles \cite{CastroNetoRMP09}. Graphene also allows
tuning the Fermi level continuously through zero energy by the
ambipolar electric field effect, and this tunability has been
intensively exploited in many experiments. In particular, it
was found that electric field strongly affects electron-phonon
interactions
\cite{PisanaNM07,YanPRL08,MalardPRL08,AndoJPSJ06,CastroNetoGuineaPRB07,AndoJPSJ07,YanPRB09},
which is hard or impossible to achieve in bulk materials.
Particularly intriguing is the case of bilayer graphene, where
the electric field breaks down the interlayer symmetry
\cite{CastroNetoRMP09} and changes qualitatively the
electron-phonon coupling \cite{AndoJPSJ09}. Up to now the vast
majority of the optical phonon studies in graphene were
performed by Raman spectroscopy, whereas infrared techniques
that can provide important complementary information, due to
different selection rules, remain largely unexplored. Here we
present an infrared study of the in-plane optical phonon mode
in gated bilayer graphene. We observe several new phonon
anomalies, of which the most spectacular one is a giant
enhancement of the phonon resonance as a function of the gate
voltage. In addition, the phonon peak is found to have a
pronounced Fano shape \cite{FanoPR61}, which indicates a
coupling of this mode to a continuum of electron-hole
excitations. We believe that this gate-tunable coupling can be
used in various electro-optical applications.

\begin{figure}[htb]
   \centerline{\includegraphics[width=7.5cm,clip=true]{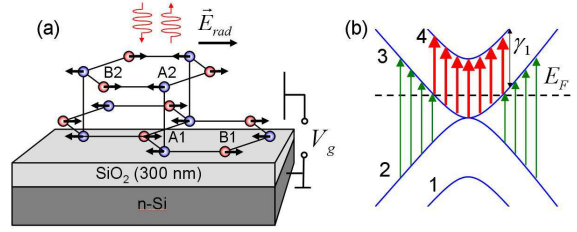}}
   \caption{(Color online) (a) A sketch of the infrared reflectivity experiment and the
   crystal structure of Bernal-stacked bilayer graphene. Atomic displacements corresponding to the antisymmetric
   mode $E_u$ are indicated by arrows; for the symmetric mode $E_{g}$ the displacements in one of the layers should be inverted.
   (b) A simplified band structure of bilayer graphene. Arrows indicate electronic transitions discussed in the text.}
   \label{FigStruct}
\end{figure}

Reflectivity spectra of an exfoliated bottom-gated bilayer
graphene flake on top of a SiO$_2$(300 nm)/Si substrate
(Fig.1a) were taken at 10 K at near-normal incidence using an
infrared microscope attached to a Fourier transform
spectrometer. Lithographically made leads allowed applying the
gate voltage $V_{g}$ and measuring resistivity during the
optical experiment. The charge neutral state corresponding to
the resistivity maximum was observed at $V_{g0}\approx$ -30 V.
Fig. 2a shows the reflectivity normalized to the bare substrate
at various gate voltages from -100 V to 100 V. The doping level
can be deduced using the relation $n$[cm$^{-2}$] =
7.2$\cdot$10$^{10}$ $\times (V_g - V_{g0})$ [V]. One can
clearly see a phonon structure at about 1600 cm$^{-1}$ (0.2 eV)
corresponding to the infrared active antisymmetric $E_{u}$ mode
(Fig.1a). In order to easier compare the phonon feature at
different values of $V_{g}$, in Fig.2b we subtract from the
reflectivity curves a smooth baseline obtained by their
low-order polynomial fitting outside the region 1500 - 1700
cm$^{-1}$. While the structure is weak in the vicinity of $V_g
= V_{g0}$, it grows quickly at elevated (both positive and
negative) gate voltages.

\begin{figure}[htb]
   \centerline{\includegraphics[width=7cm,clip=true]{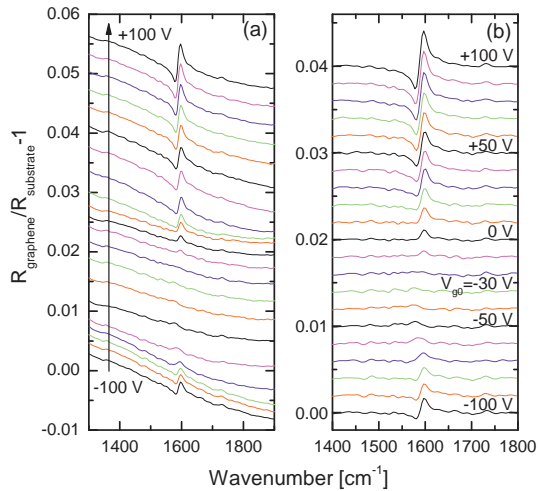}}
   \caption{(Color online)(a) infrared reflectivity spectra of bilayer graphene on SiO$_{2}$(300
   nm)/Si around the $E_{u}$ peak normalized by the reflectivity of the bare
   substrate at different gate voltages. (b) the same spectra with a baseline subtracted.
   The curves (except at -100 V) are vertically shifted.}
   \label{FigRefl}
\end{figure}

It is not straightforward to interpret the phonon lineshapes in
the raw reflectivity spectra, which are sensitive to the choice
of the substrate. In Fig.3 we show the real part of the optical
conductivity of graphene, $\Delta\sigma_{1}(\omega)$, extracted
using a Kramers-Kronig consistent variational spectra analysis
\cite{KuzmenkoRSI05} that takes into account the known optical
constants and thicknesses of the SiO$_{2}$ and Si layers. The
smooth background due to direct optical absorption by
electronic transitions, which is discussed in detail elsewhere
\cite{KuzmenkoPRB09,KuzmenkoToBePublished}, is subtracted.

A few interesting observations can be made. First of all, the
peak intensity grows rapidly as the gate voltage is tuned away
from $V_{g0}$. Second, the phonon peak is clearly asymmetric,
with a characteristic Fano lineshape \cite{FanoPR61} that is
most pronounced at not very high values of $|V_g - V_{g0}|$
(for example, at $V_{g}$ = -70, -80, 0, +10 V). Third, close to
the charge neutral point (for example at -40 and -50 V), the
peak is {\em negative}, i.e. the conductivity has a weak but a
distinct dip. Fourth, the position of the peak maximum is
gate-voltage dependent. For example, one can see that for
$V_{g}> 40$ V it shifts to lower frequencies as the gate
voltage is increased.

\begin{figure}[htb]
   \centerline{\includegraphics[width=8cm,clip=true]{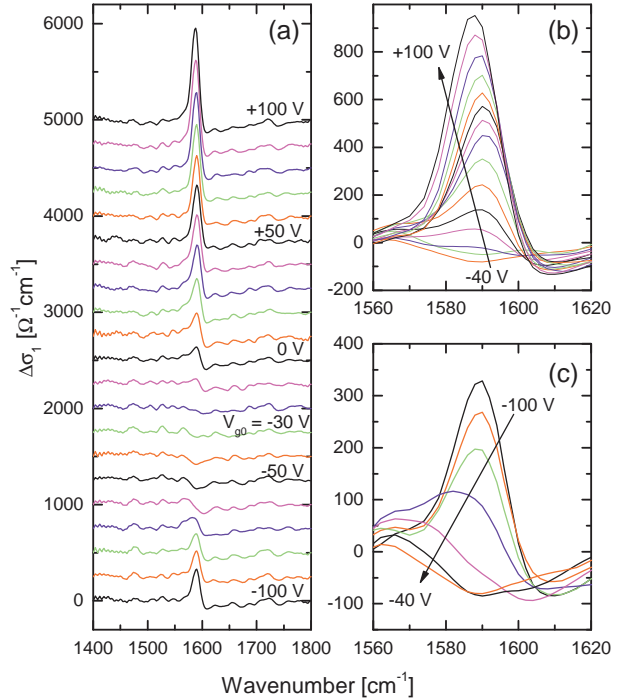}}
   \caption{(Color online)(a) The real part of the optical conductivity $\Delta\sigma_{1}(\omega)$ of bilayer
   graphene at different gate voltages. The electronic baseline is subtracted. The curves (except at -100 V) are vertically shifted.
   (b) and (c): the expanded views of the $E_u$ peak at electron and hole doping.}
   \label{FigS1}
\end{figure}

The Fano lineshape provides evidence of a coupling between this
sharp mode and a continuum of electronic excitations in the
same energy range. Following the general Fano
theory\cite{FanoPR61}, the optical conductivity
$\Delta\sigma_{1}(\omega)$ of such coupled system minus the
background due to the bare continuum can described by the
formula:
\begin{equation}\label{Fano}
\Delta\sigma_{1}(\omega)=
\frac{\omega_{p}^2}{4\pi\Gamma}\frac{q^{2} + 2 q z - 1}{q^2(1 +
z^2) }
\end{equation}
\noindent with $z = 2(\omega - \omega_{0})/\Gamma$. Here
$\omega_{0}$ is the phonon frequency, $\Gamma$ is the linewidth
and $\omega_{p}$ is the plasma frequency, which is proportional
to the infrared effective charge as specified below. The
asymmetry is described by a dimensionless parameter $q$ so that
the symmetric Lorenzian lineshape is recovered in the limit
$|q|\rightarrow\infty$. The doping dependence of the Fano
parameters extracted by the fitting of the conductivity spectra
of Fig.3 is shown in Fig.4a-d.

\begin{figure}[htb]
   \centerline{\includegraphics[width=7.5cm,clip=true]{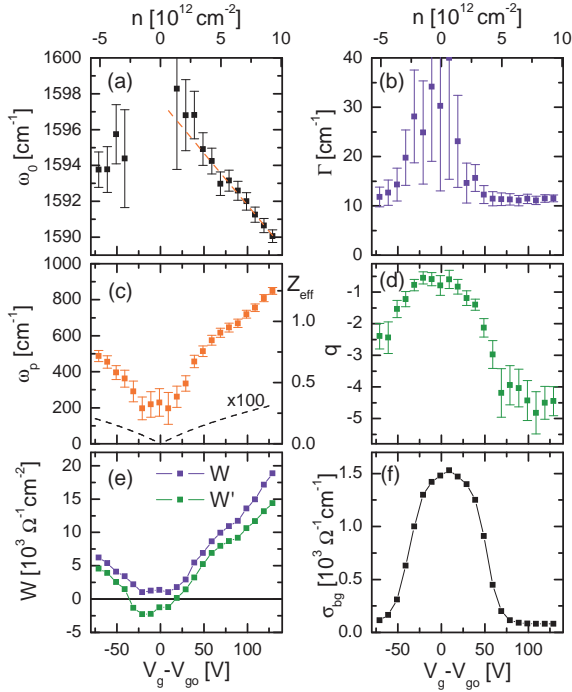}}
   \caption{(Color online)(a)-(d) Fano parameters $\omega_{0}$, $\Gamma$, $\omega_{p}$ ($Z_{\mbox{\scriptsize eff}}$) and $q$
   of the phonon peak as a function of doping. At very low doping,
   the error bars for $\omega_{0}$ exceed the panel scale.
   The effective charge corresponding to the calculated static dipole is shown by the dashed line in
   panel (c). (e) The bare spectral phonon weight $W$ and the peak
   area $W'$. (f) The background electronic conductivity at the
   phonon energy.}
   \label{FigFano}
\end{figure}

From Fig.4a one can see that $\omega_{0}$ decreases almost
linearly with the slope $\partial\omega_{0}/\partial V_{g}
\approx$ -0.058 cm$^{-1}$/V. Note that due to the Fano shape,
$\omega_{0}$ is {\em larger} than the position of the
conductivity peak maximum, the difference being dependent on
the asymmetry parameter. This is why the softening is seemingly
less pronounced in Fig.\ref{FigS1}b and \ref{FigS1}c. The
linewidth is enhanced at low doping levels and decreases with
the increasing the gate voltage. The fitted value of $\Gamma$
shows a saturation at 10 cm$^{-1}$, which is the resolution
limit of this measurement and therefore an overestimation of
the true linewidth. The plasma frequency $\omega_{p}$ is at
minimum close to $V_{g0}$ (Fig.4c), where $\omega_{p} \sim 200$
cm$^{-1}$ and strongly increases with the gate voltage. In
Fig.4e we plot the bare phonon spectral weight $W =
\omega_{p}^2/8$ as well as the peak area $W' \equiv \int
\Delta\sigma_1(\omega)d\omega$ (integrated between 1570 and
1610 cm$^{-1}$). Note that in the presence of coupling $W'
\approx \omega_{p}^{2}(1 - 1/q^2)/8\neq W$. If the asymmetry is
very strong ($|q| < 1$) then the area becomes negative. This is
because $W'$ is affected not only by the direct optical
absorption due to the phonon mode but also by the
coupling-induced transformations of the optical response of the
continuum. In particular the negative value of $W'$ is
explained by the removal of spectral weight from the continuum.
The asymmetry parameter $|q|$ also shows a strong doping
dependence, with a decrease as the doping is increased. This
decrease is not monotonic: the strongest changes are observed
at $|V_{g} - V_{g0}| \approx$ 50 V, while at higher doping the
asymmetry remains roughly constant.

Let us discuss the observed phonon anomalies and their relation
to the band structure and electronic excitations. The frequency
shift $\Delta\omega_{0}$ and the linewidth $\Gamma$ are related
to the phonon self energy $\Pi(\omega)$: $\Delta\omega_{0}
\approx \hbar^{-1}\mbox{Re } \Pi(\omega_{0})$, $\Gamma \approx
-\hbar^{-1}\mbox{Im } \Pi(\omega_{0})$. The self energy for the
zone-center optical phonons is given at zero temperature by a
sum of the contributions of all vertical (momentum conserving)
transitions across the Fermi level weighted with their
respective electron-phonon matrix elements. Each contribution
has a Lorentzian spectral shape $(\Delta\epsilon - \hbar\omega
+ i\delta)^{-1}$, where $\Delta\epsilon$ is the transition
energy and $\delta$ is a broadening parameter. Therefore
$\Gamma$ is determined only by the transitions that overlap
with the phonon energy within $\delta$, while
$\Delta\omega_{0}$ is affected by all transitions with non-zero
electron-phonon matrix elements.

A simplified band structure of bilayer graphene, where we
neglect the electron-hole asymmetry and the bandgap, is shown
in Fig.1b. Each of the hole and electron bands is split into
two parallel subbands (numbered 1,2 and 3,4 respectively)
separated by the interplane hopping energy $\gamma_{1} \sim$
0.4 eV. We shall refer to the case of electron doping (positive
gate voltages) although the results are equally valid for hole
doping. Based on this band structure, Ando calculated the
phonon self energy and predicted the softening of the $E_{u}$
phonon with doping \cite{AndoJPSJ07}, in an excellent agreement
with the present infrared measurement. Interestingly, the
softening was also found in a Raman study \cite{MalardPRL08},
where the $E_{u}$ mode was observable due to a possible
breaking of the inversion symmetry of bilayer graphene in the
used device. By comparing the observed slope
$\partial\omega_{0}/\partial V_{g}$ with the theoretical curve
we estimate the dimensionless electron-phonon coupling
parameter $\lambda$ defined in Ref.\onlinecite{AndoJPSJ07} to
be about $6 \times 10^{-3}$, in agreement with the Raman study
\cite{MalardPRL08}. According to Ando \cite{AndoJPSJ07}, the
softening of the $E_{u}$ mode is due to the contribution to the
self energy coming from the transitions between the bands 3 and
4  (thick red arrows in Fig.1b). Since the two bands are
parallel, these transitions form a narrow peak centered at
$\gamma_{1}$. Its intensity grows rapidly with doping in the
course of the progressive filling of the band 3.

It is important that none of the interband transitions, except
the ones between the bands 2 and 3, overlap with the phonon
energy, as $\gamma_1$ is about twice as large as $\omega_{0}$.
On the other hand, the electron-phonon matrix elements between
the $E_{u}$ mode and the $2\rightarrow3$ excitations vanish
according to the symmetry analysis \cite{AndoJPSJ07}.
Therefore, the same calculation \cite{AndoJPSJ07} predicts a
very small and weakly doping dependent linewidth of the $E_{u}$
phonon mode. Instead, we observe that $\Gamma$ is large ($\sim$
20-30 cm$^{-1}$) at low doping and strongly decreases with $n$
(Fig.4b).

We argue that the large linewidth and the Fano shape of the
phonon peak are likely due to the interaction with the
$2\rightarrow3$ transitions (shown with thin green arrows in
Fig.1b), despite the fact that the self-energy calculation
Ref.\cite{AndoJPSJ07} would suggest a negligible coupling in
this case.

The $2\rightarrow3$ transitions form a continuum of
particle-hole excitations. Such a continuum is also present in
monolayer graphene and in graphite, where it results in the
universal optical conductivity of $e^2/4\hbar$
\cite{AndoJPSJ02,KuzmenkoPRL08,NairScience08,LiNP08}. The
continuum starts at $\omega = 2E_F$ due to the Pauli principle
\cite{AndoJPSJ02}. If $E_F < \omega_{0}/2$ then the phonon
frequency lies {\em inside} the $2\rightarrow3$ continuum and
in this case one expects the linewidth and the asymmetry to be
stronger. Indeed, we see a clear correlation between the sudden
decrease of the asymmetry parameter $q$ (Fig.4d), the decrease
of the broadening (Fig.4b) and the reduction of the background
electronic optical conductivity at the phonon frequency
$\sigma_{bg}(\omega_{0})$ (Fig.4e)
\cite{KuzmenkoToBePublished}, which corresponds to $E_F \approx
\omega_{0}/2$. It is worth mentioning that the anomalous
softening of the Raman-active phonon mode in bilayer graphene
is observed at about the same gate voltage \cite{YanPRL08}.

Among the possible reasons of the "forbidden" interaction
between the $E_{u}$ mode and the $2\rightarrow3$ electronic
excitations we can mention (i) the electron-hole asymmetry,
which is clearly observed in the electronic infrared spectra
\cite{ZhangPRB08,LiPRL09,KuzmenkoPRB09,KuzmenkoToBePublished}
and (ii) an electric field generated by the gate electrode and
charged impurities which breaks the inversion symmetry
\cite{AndoJPSJ09}. Note that a Fano phonon lineshape in the
infrared spectra of bilayer graphene was also observed in
Ref.\cite{ZhangNature09}.

Finally, we discuss the most spectacular effect of this study:
a large increase of the phonon intensity as a function of
doping. From the plasma frequency we can deduce the
dimensionless effective infrared charge of the mode
$Z_{\mbox{\scriptsize eff}} = \omega_{p}(m_{c}V/(4\pi N
e^2))^{1/2}$, where $N = 4$ is the number of atoms in the unit
cell, $V$ is the unit cell volume, $m_{c}$ is the mass of the
carbon atom and $e$ is the elementary charge. The doping
dependence of the effective charge is shown in Fig.4c. One can
see that at high gate voltages $Z_{\mbox{\scriptsize eff}}$
exceeds 1, which is the value observed in entirely ionic
compounds like NaCl and LiF. We first compare this value to the
effective charge of the static dipole created because the doped
charge is distributed not equally between the atomic positions
A1(A2) and B1(B2) (Fig.1a). The static charge calculated using the
tight-binding model (the dashed line in Fig.4c) appears to be
more than 300 times smaller than the observed infrared
(dynamical) effective charge. Such an utter discrepancy calls
for a radically different physical approach.

Notably, similar effects of a large enhancement of vibrational
resonances were observed in organic linear-chain conductors
such as K-TCNQ \cite{TannerPRB77} and in doped C$_{60}$
\cite{FuPRB92}. Effective infrared charges significantly larger
than ionic charges were also found in some transition metal
compounds, for example in FeSi \cite{DamascelliPRB97}. These
findings were explained using the so-called "charged-phonon"
theory of M.J.Rice et al. \cite{RicePRL77}. Within this model,
the molecular vibrations that are usually not active or weakly
active in the infrared range become enhanced due to a coupling
to electronic resonances at higher energy (such as the
transition $t_{1u}\rightarrow t_{1g}$ in doped $C_{60}$). Their
effective charge is effectively "borrowed" from the electronic
transitions, in accordance with the f-sum rule.

A clear correlation between the intensity of the phonon
resonance and the strength of the electronic transition was
experimentally observed \cite{FuPRB92}. The charged-phonon
model thus seems to be relevant to the intensity enhancement of
the $E_{u}$ mode in bilayer graphene, in view of the
observation that the strong doping dependence of the intensity
of the 3$\rightarrow$4 transition correlates with the growth of
the phonon peak. Interestingly, a simultaneous increase of the
phonon intensity and electronic absorption with doping was also
found in organic field effect transistors
\cite{LiNanoLetters06}. We note that even at zero doping, where
the intensity of the $3\rightarrow 4$ transition is zero, the
effective charge is about 0.25. This suggests that other
transitions ($1\rightarrow3$, $2\rightarrow4$ and
$1\rightarrow4$), which remain populated at all doping levels,
also contribute to $Z_{\mbox{\scriptsize eff}}$. We notice that
the zero-doping value of $Z_{\mbox{\scriptsize eff}}$ is
comparable with the ones reported in graphite
\cite{NemanichSSC77}.

As a concluding remark, the possibility to control the phonon
intensity by the gate voltage might be of practical importance,
for example, in optoelectronics. We also suggest that the peak
can be used as a doping indicator.

In summary, we observed several anomalies of the infrared
active $E_{u}$ mode in bilayer graphene: (i) a drastic
intensity enhancement and a softening as a function of doping,
(ii) a strong Fano-like asymmetry of the lineshape, with even a
{\em negative} peak area at low doping. We suggest that the
coupling to the narrow transition $3\rightarrow 4$ is
responsible for the mode softening \cite{AndoJPSJ07} and the
intensity enhancement, via the charged-phonon mechanism
\cite{RicePRL77}. The interaction with the continuum
$2\rightarrow 3$ likely gives rise to the asymmetry and the
broadening of the phonon peak. Further studies are required to
understand quantitatively the complicated picture of the
electron-phonon coupling in graphene. This work was supported
by the Swiss National Science Foundation (SNSF) by the grant
200021-120347, through the National Center of Competence in
Research "Materials with Novel Electronic Properties-MaNEP" and
by Italian MIUR project PRIN 2007FW3MJX.

\end{document}